\documentstyle[12pt]{article}

\begin{document}

\title{Classical Oscilators in General Relativity
\footnote{Published in Europhysics Letters, {\bf 49}:(2) 156-161 JAN, 2000}}

\author{Ion I. Cot\u aescu \\ and\\Dumitru N. Vulcanov\\ {\small \it The West 
                 University of Timi\c soara,}\\
   {\small \it V. P\^ arvan Ave. 4, RO-1900 Timi\c soara, Romania}}

\date{}
\maketitle

\begin{abstract}
It is shown that in general relativity some static metrics are able to 
simulate oscillatory motions. Their form depends on two 
arbitrary real parameters which determine the specific oscillation modes. 
The conclusion is that these metrics  can be used for new geometric models of  
closed  or even open bags.      

Pacs: 04.02.-q and 04.02.Jb

\end{abstract}
\

\newpage

\section{Introduction}
\

In general relativity the quark confinement can be treated as a pure geometric 
effect produced by  an anti-de Sitter cavity \cite{P1,P2}. This model is based 
on a (3+1)-dimensional anti-de Sitter static metric  \cite{P2} which  
reproduces the usual equations of motion  of the nonrelativistic (isotropic) 
harmonic oscillator 
(NRHO). On the other hand, it is known that the quantum massive scalar field on 
anti-de Sitter spacetime  has an equidistant discrete energy spectrum \cite{O}. 
Thus, the classical or quantum particle  inside of an anti-de Sitter bag 
appears as the natural relativistic correspondent of the classical or 
quantum NRHO.  

However, we can ask if there are other metrics that may be  used for new 
geometric  
bag models. In our opinion, these must  produce similar kinetic effects as the 
anti-de Sitter one, namely closed trajectories inside of the bag. Obviously, it 
is very difficult to classify all the metrics with this property but if  we 
restrict ourselves to the simple particular case of the oscillatory motions, 
arising  from linear equations, then we could find some new interesting 
metrics. This is the motive why  we would like to study here  metrics 
producing  timelike geodesics  which (I) are linear in  space Cartesian 
coordinates  and their time derivatives, and (II)  have the NRHO as 
nonrelativistic limit. These systems will be called generally relativistic 
oscillators (RO). Our objective is to find a representative class of 
(3+1) metrics satisfying  these two requirements and to analyze the 
geodesic motion of these models, in order to point out new possible 
relativistic effects. 

We show that there are two interesting types of metrics with these properties. 
The first one contains deformed or conformaly transformed anti-de Sitter 
metrics. These describe  spherical closed cavities where the confined particles 
move as harmonic or rotating oscillators. Another type is that of the deformed 
or conformaly transformed de Sitter metrics of open bags. In these models the 
particle remains inside of the bag only for energies less than a given limit. 
If the energy increases over this limit  then the particle escapes to infinity.  

We start in Sec.2  with a short review of the properties of the relativistic 
central motion by taking into account the conservation of the energy and 
angular momentum. In the next section we introduce a family of RO accomplishing 
(I) and (II), the  metrics  of which depend on two real parameters. The  
classical  motion of these RO is studied in Sec.4 where we obtain the mentioned 
results.

\section{The relativistic central motion}
\

In general relativity the  correspondent of the nonrelativistic  classical 
central motion is the geodesic  motion on  central (i.e., spherically symmetric)  
static charts where the line element in holonomic Cartesian coordinates, 
$x^{0}=ct$ and $x^{i}, i=1, 2, 3$, is invariant under time translations and 
the space rotations, $R\in SO(3)$, of the space coordinates, 
$x^{i}\to R^{i}_{j}x^{j}$. The most general form of the line element is
\begin{equation}\label{(hhh)} 
ds^{2}=A(r)dt^{2}-[B(r)\delta_{ij}+C(r)x^{i}x^{j}]dx^{i}dx^{j}\,.
\end{equation} 
where $A$, $B$ and $C$ are arbitrary differentiable functions of $r=|{\bf x}|$.

The geodesic equations of a test particle of mass $m$ can be derived 
directly in terms of $x^{i}(t)$ and their time derivatives, $\dot{x^{i}}$ and 
$\ddot{x_{i}}$, by  using  the  Hamiltonian formalism. Moreover, this  help us 
to write down the conservation laws given by the Noether theorem. Thus we 
obtain the conservation  of the  energy 
\begin{equation}\label{(ee)}
E= mc\,A\frac{dt}{ds}\,,
\end{equation}
and of the angular momentum
\begin{equation}\label{(ll)}
L^{ij}=E\frac{B}{A}(x^{i}\dot{x^{j}} -x^{j}\dot{x^{i}})\,.
\end{equation}
With these ingredients, after a few manipulations we find the equations of the 
timelike geodesics, 
\begin{equation}\label{(carteq)}
\ddot x^{i}+\frac{1}{r} \left(\frac{A}{B}\right)'\frac{L^{ij}}{E}\dot{x^{j}}
+\partial_{i}W=0\,, 
\end{equation}
written with the notation $'=\partial_{r}$. Here 
\begin{equation}\label{(cww)}
W=\frac{1}{2}\frac{A}{B+r^{2}C}\left(\frac{m^{2}c^2}{E^{2}}A-1-
\frac{L^{2}}{E^{2}}\frac{AC}{B^{2}}\right)
\end{equation}
is a function of $r$ playing the role of potential. These equations explicitly 
depends on $E$ and $L^{ij}$ or $L=|{\bf L}|$, used here as integration constants. 
Thus we obtain  an alternative  form of the well-known  geodesics  of the 
relativistic central motion \cite{F,W} from which one can easily recover their 
main properties. For example, we see that the trajectories are  
on  space shells orthogonal to ${\bf L}$. In other respects, we recognize that  
the second term of (\ref{(carteq)}) is just the Coriolis-like relativistic 
contribution \cite{LL}.   

The form of the potential (\ref{(cww)}) is complicated since it is strongly 
non-linear in $A$, $B$ and $C$. For this reason it is convenient to perform a 
change of functions considering three new functions of $r$, denoted by 
$\alpha$, $\beta$ and $\gamma$, such that
\begin{equation}\label{(ABC)}   
A=c^{2}\frac{\alpha}{\beta}, \quad B=\frac{\alpha}{\beta \gamma}, \quad 
C= \frac{\alpha}{\beta^{2} \gamma}\frac{\gamma-\beta}{r^2}\,.
\end{equation}
Then  Eqs.(\ref{(carteq)}) become  
\begin{equation}\label{(carteq1)}
\ddot x^{i}+\frac{\gamma'}{r} \frac{L^{ij}}{E}\dot x^{j}
+\partial_{i}W=0\,, 
\end{equation}
where 
\begin{equation}\label{(20)}
W=\frac{c^2}{2}\left( \alpha\frac{m^{2}c^4}{E^2}-\beta+
\gamma(\beta-\gamma)\frac{1}{r^2} \frac{L^{2}c^2}{E^2} \right)
\end{equation}
has  a more comprehensible form.

The geodesic equations in spherical coordinates can be obtained directly in 
terms of  functions $\alpha$, $\beta$ and $\gamma$ starting with the line 
element   
\begin{equation}\label{(mesf)}
ds^{2}=c^{2}\frac{\alpha}{\beta}dt^{2}-\frac{\alpha}{\beta^2}dr^{2}-
\frac{\alpha}{\beta \gamma}r^{2}(d\theta^{2}+\sin^{2}\theta d\phi^{2})\,.
\end{equation}
If we consider  the conservation of the energy and that of the 
angular momentum chosen along to the 3-axis, we obtain the geodesic equations  
on the space shell $\theta=\pi/2$, 
\begin{eqnarray}
\ddot r+w'=0, \label{(100)}\\ 
\dot \phi=\frac{\gamma}{r^2}\frac{Lc^2}{E},\label{(110)}
\end{eqnarray}
where
\begin{equation}\label{(120)}
w=\frac{c^2}{2}\left( \alpha \frac{m^{2}c^4}{E^{2}}-\beta + 
\frac{\beta\gamma}{r^2}\frac{L^{2}c^2}{E^2} \right)\,.
\end{equation}
Notice that the prime integral ${\dot r}^{2}+2w=0$ allows us to directly 
integrate the radial equation (\ref{(100)}) as in nonrelativistic mechanics. 

Thus we obtain  new forms of the geodesic equations of the relativistic 
central motion  that have some technical advantages. One of them is that the 
potentials $W$ or $w$ are linear in $\alpha$ and $\beta$ ia a such a manner that
the potential energy will have a term proportional with $\alpha-\beta$ in the 
nonrelativistic limit. In addition, from (\ref{(carteq1)}) we see that the 
Coriolis-like relativistic effect is due  only to the first derivative of the 
function $\gamma$. Therefore, this effect is canceled  when $\gamma$ is a 
constant.

\section{The metrics of relativistic oscillators}
\

Let us turn now to the problem of our RO which must satisfy  the conditions (I) 
and (II). We shall show that these can be accomplished if we consider the 
particular functions   
\begin{equation}\label{(abc)}
\alpha_{o}=1+(1+\lambda){\hat\omega}^{2} r^{2}, \quad 
\beta_{o}=1+\lambda{\hat\omega}^{2} r^{2}, \quad
\gamma_{o}=1+\lambda'{\hat\omega}^{2} r^{2},
\end{equation}
where $\hat \omega=\omega/c$ and $\lambda$ and $\lambda'$ are two arbitrary 
real parameters. Indeed, if we take  $L=L^{12}\not=0$ and $L^{23}=L^{13}=0$ in 
order  to keep the trajectory  on the shell $x^{3}=0$, then Eqs.(\ref{(carteq1)}) 
take the form 
\begin{eqnarray}\label{(2eq)}
\ddot{x^{1}}+2\Omega_{c} \dot{x^{2}}+{\Omega_{0}}^{2}x^{1}=0\,,\nonumber\\
\ddot{x^{2}}-2\Omega_{c} \dot{x^{1}}+{\Omega_{0}}^{2}x^{2}=0\,,
\label{(eq1)}
\end{eqnarray}
which is linear in $x^i$ and their time derivatives. The constant 
coefficients  herein are the Cartesian effective frequency squared,
\begin{equation}\label{(200)}
{\Omega_{0}}^{2}=\omega^{2}\left[(1+\lambda)\frac{m^{2}c^4}{E^{2}}-\lambda+
\lambda'(\lambda-\lambda')\frac{L^{2}\omega^2}{E^{2}}\right]\,,
\end{equation}
and the Coriolis-like frequency  
\begin{equation}\label{(210)}
\Omega_{c}=\lambda'\frac{L\omega^2}{E}\,.
\end{equation}
On the other hand, we observe that in the nonrelativistic limit, 
for  small values of the nonrelativistic energy $E_{nr}=E-mc^{2}$ and 
$c \to \infty$, we have
\begin{equation}\label{(lim)}
\lim_{c\to\infty}\Omega_{0}=\omega\,, \qquad
\lim_{c\to\infty}\Omega_{c}=0\,,
\end{equation}
which means that all these RO lead to NRHO. Notice that in the absence of the 
interaction ($\omega\to 0$) the metric becomes a flat one.

Thus we have found a family of metrics which satisfy (I) 
and (II). According to  (\ref{(hhh)}), (\ref{(ABC)}) and (\ref{(abc)}), these 
metrics are given by the line elements in Cartesian coordinates 
\begin{equation}\label{(mf)}
ds^{2}=c^{2}\frac{\alpha_{o}}{\beta_{o}}dt^{2} - \frac{\alpha_{o}}{ 
\beta_{o}\gamma_{o}} \left[\delta_{ij}-(\lambda-\lambda')\frac{\hat \omega^{2}}
{\beta_{o}} 
x^{i}x^{j}\right]dx^{i}dx^{j}\,, 
\end{equation}
while in spherical coordinates the line elements can be obtained directly 
by replacing  (\ref{(abc)}) in (\ref{(mesf)}). 

These metrics  represent in some sense a  generalization 
of the anti-de Sitter one. In order to avoid supplementary singularities  
we restrict ourselves to  $\lambda'\ge 0$. Then we see that 
the metrics with $\lambda<0$  are either deformations, if $\lambda'\not=0$, or 
conformal transformations, if $\lambda'=0$, of some anti-de Sitter metrics. 
These are  are singular on the sphere of the radius 
$r_{e}=c/\omega\sqrt{-\lambda}$ which is just the event horizon of an observer 
situated at $x^{i}=0$. Particularly,  $\lambda=-1$ and $\lambda'=0$ 
correspond to the  anti-de Sitter metric \cite{P2}. 
The metrics with $\lambda>0$  and  $\lambda'\not=0$ are 
deformations of some de Sitter static metrics while for $\lambda'=0$ 
these are just conformal transformations of the de Sitter one.  
However, the  de Sitter metric is not included in this family since 
it is not able to produce oscillations. 

In general, the metrics we have introduced can be considered as solutions of 
the Einstein equations with sources but only in the presence of the cosmological 
term which  must give rise to the anti-de Sitter metric when the space is 
devoid of matter. Our preliminary calculations indicate that, by choosing 
suitable values of  cosmological constant, one can find large domains of 
parameters where the stress energy tensor of the gravitational sources  
satisfies at least the dominant energy condition \cite{R1,R2}.   

\section{Trajectories and oscillation modes}
\

The values of the parameters $\lambda$ and 
$\lambda'$   determine the oscillation modes of  RO. 
In Cartesian coordinates these  appear as  two orthogonal  harmonic 
oscillators  coupled through velocities because of the Coriolis-like term. 
One can verify that they have two oscillation modes with the frequencies 
$\Omega_{\pm}=\Omega\pm \Omega_{c}$. Hence it results that our RO are not  
isotropic harmonic oscillators if $\Omega_{c}\not= 0$.

However, these oscillation modes can be better analyzed  in spherical 
coordinates. From (\ref{(100)})-(\ref{(120)}) and (\ref{(abc)}) we have    
\begin{eqnarray}
&&\ddot r+{\Omega}^{2}r -\frac{L^{2}c^{4}}{E^{2}}\frac{1}{r^3}=0\label{(eq2)},\\
&&\dot \phi = \frac{Lc^2}{E}\frac{1}{r^2}+\Omega_{c},\label{(eq3)}
\end{eqnarray} 
where
\begin{equation}\label{(220)}
{\Omega}^{2}={\Omega_{0}}^{2}+{\Omega_{c}}^{2}
=\omega^{2}\left[(1+\lambda)\frac{m^{2}c^4}{E^{2}}-\lambda+
\lambda\lambda'\frac{L^{2}\omega^2}{E^{2}}\right]
\end{equation}
is the radial frequency squared. These equations can be easily integrated obtaining 
the solutions  
\begin{eqnarray}
r(t)&=&\frac{1}{\sqrt{\Omega}}[\kappa _{1}+\kappa_{2}\sin 2(\Omega t+
\delta_{1})]^{1/2}\label{(er)},\\    
\phi (t)&=& \delta _{2} +\Omega _{c}t+  \arctan 
\left\{ \frac{E}{Lc^2}[\kappa_{1} \tan(\Omega t+\delta_{1})+\kappa_{2}] 
\right\},\label{(epsi)}
\end{eqnarray}
where
\begin{equation}\label{(kap1)}
\kappa_{1}=\frac{c^2}{2\Omega}\left[ 1-\frac{m^{2}c^4}{E^2}-(\lambda+\lambda')
\frac{L^{2}\omega^2}{E^2}\right]
\end{equation}
and
\begin{equation}\label{(kapa12)}
{\kappa_{1}}^{2}-{\kappa_{2}}^{2}=\frac{L^{2}c^4}{E^2}\,.
\end{equation}
The phases $\delta_{1}$ and $\delta_{2}$ are the remaining integration 
constants after we have fixed the values of the energy and of the three 
components of the angular momentum. These phases are determined by the initial 
conditions (at $t=0$)
\begin{equation}
r(0)=r_{0}\,, \qquad \phi (0)=\phi_{0}\,.
\end{equation}

The Eqs.(\ref{(er)}) and (\ref{(epsi)}) are similar 
 to those of an isotropic harmonic oscillator of the frequency $\Omega$ apart  
the second term of Eq.(\ref{(epsi)}) which produces an uniform 
rotation of the angular velocity $\Omega_{c}$. This means that the RO with 
$\lambda'\not=0$ are in fact relativistic {\it rotating oscillators}. However, 
when $\lambda'=0$ we have $\Omega_{c}=0$ and, consequently, the RO becomes an 
harmonic oscillator. If, in addition, we take $\lambda=-1$ then we obtain the 
anti-de Sitter oscillator of Ref.\cite{P2}. 

According to (\ref{(kapa12)}) the motion is possible only if 
\begin{equation}
 \kappa_{1}\ge \frac{Lc^2}{E}\,.
\end{equation}
This condition combined with (\ref{(kap1)}) gives us the 
energy spectrum for a fixed $L$ as 
\begin{equation}
E^{2}\ge (mc^{2}+L\omega)^{2}-(1+\lambda-\lambda')L^{2}\omega^{2}\,.
\end{equation}
On the other hand, we observe that for  $\lambda>0$  there are  values of $E$ 
and $L$  for which the oscillatory motion  degenerates into  open (uniform 
or accelerated) motions. This is because the oscillations appear only when 
\begin{equation} 
\Omega^{2}>0
\end{equation}
which, according to (\ref{(220)}), gives the oscillation condition     
\begin{equation}\label{(con)}
\lambda E^{2}< (1+\lambda)m^{2}c^{4} 
+\lambda\lambda'L^{2}\omega^{2}\,.
\end{equation}  
When  both these conditions are accomplished we have an oscillatory motion
with a trajectory of an ellipsoidal form on the ring  
$r\in [r_{min},r_{max}]$ where
\begin{equation}
r_{min}=\sqrt{\frac{\kappa_{1}-\kappa_{2}}{\Omega}}\,, \qquad
r_{max}=\sqrt{\frac{\kappa_{1}+\kappa_{2}}{\Omega}}\,.
\end{equation}
Particularly, for $L=0$ we obtain the same result as in the case of 
(1+1) RO, namely  that  the particles of the 
models  with $\lambda>0$ oscillate only if  $mc^{2}<E<mc^{2}\sqrt{1+1/\lambda}$ 
while those having $\lambda\le 0$ oscillate for all the possible energies, 
$E>mc^2$ \cite{COTA1}.   

What is interesting here is that for all  $\lambda\not=-1$ and 
$\lambda'\not=0$ the 
frequency $\Omega$  depends on $E$ and $L$. This can be interpreted as a pure 
relativistic effect since in the 
nonrelativistic limit $\Omega\to \omega$, as it results from (\ref{(lim)}).
Moreover, then we have
\begin{equation}
\kappa_{1}\to \frac{E_{nr}}{\omega m}, \quad \kappa_{2}\to \frac{1}{\omega m}
\sqrt{{E_{nr}}^{2}-L^{2}\omega^{2}}
\end{equation}
such that the trajectory becomes just the well-known nonrelativistic one.
Notice that in the particular case of the harmonic oscillators (with 
$\lambda'=0$) the frequencies   depend only on 
$E$ like in the case of the (1+1) RO \cite{COTA1}.

\section{Conclusions}
\

Here we have defined a new family of RO as a possible generalization  
of the anti-de Sitter oscillator. The RO with $\lambda<0$ are models of 
closed bags since their metrics  are singular on the sphere of the radius 
$r_{e}$ and, therefore, the oscillating particle remains confined to 
cavity interior. We specify that our new parameter $\lambda$  can be seen as a 
supplementary fit parameter of the cavity radius. The metrics with $\lambda>0$  
have no singularities and, therefore, the free 
particles  can have either closed trajectories, when the motion is oscillatory 
because of the values of $E$ and $L$ satisfying the condition (\ref{(con)}), 
or open trajectories if this condition is not fulfilled. Hence these metrics 
are of new possible  models of open bag  from which the particle can escape 
when it has  enough energy. 

Thus we have an example of a  family of metrics giving oscillatory 
motions with specific kinetic effects. The next step may be to classify all the 
metrics able to generate oscillations and to establish the physical context 
in which such metrics can be produced by gravitational sources. In our 
opinion this problem is sensitive and must be carefully analyzed. We hope that 
the method presented  here should be useful for further developments in this 
direction.  
On the other hand, our family of metrics is interesting from the quantum 
point of view since all the models of quantum spinless RO with  
$\lambda'=\lambda+1$ are analytically solvable \cite{COTA2}. Thus we can 
compare the classical and quantum kinetic effects of a large set of simple 
models. This could be useful for better understanding the relation between 
the classical and  quantum approaches in general relativity.


\end{document}